\documentclass{jpp}
\pdfoutput = 1
\usepackage{graphicx}
\usepackage{natbib}

\usepackage{mathtools}
\usepackage[T1]{fontenc}
\usepackage{bm}
\usepackage{color}
\usepackage{amsmath}
\usepackage{amssymb}
\usepackage{amsfonts}
\usepackage{courier}
\usepackage{txfonts}

\newcommand{\nofrac}[2]{#1/#2}

\newcommand{\sub}[1]{\ensuremath{_{\text{#1}}}}
\newcommand{\supp}[1]{\ensuremath{^{\text{#1}}}}
\newcommand{\tightapprox}{\ensuremath{\,{\approx}\,}}
\newcommand{\tightequal}{\ensuremath{\,{=}\,}}
\newcommand{\rd}{\ensuremath{\mathrm{d}}}
\newcommand{\fe}{f\sub{e}}

\usepackage{ae}
\usepackage{units}

\newcommand{\JPPmatrix}[1]{\ensuremath{\mathsfbi{#1}}}
\renewcommand{\vec}[1]{\ensuremath{\bm{#1}}}

\usepackage{enumerate}
\usepackage[T1]{fontenc}
\usepackage{bm}
\usepackage{color}
\usepackage{graphicx}
\usepackage{hyperref}
\hypersetup{
  unicode=false,          
  pdftoolbar=true,        
  pdfmenubar=true,        
  pdffitwindow=false,     
  pdfstartview={FitH},    
 pdftitle={Evaluation of the Dreicer runaway generation rate in the presence of high-Z impurities using a neural network},    
  pdfauthor={Hesslow, Unnerfelt, Vallhagen, Embreus, Hoppe, Papp, and Fülöp},     
  pdfnewwindow=true,      
  colorlinks=true,        
  linkcolor=blue,         
  citecolor=blue,         
  filecolor=blue,         
  urlcolor=blue           
}

\usepackage{amsmath}
 \usepackage{gensymb}

\title[Runaway-electron generation rate using a neural network]{Evaluation of the Dreicer runaway generation rate in the presence of high-Z impurities using a neural network}

\author{L.~Hesslow\aff{1}\corresp{\email{hesslow@chalmers.se}}, L.~Unnerfelt\aff{1}, O.~Vallhagen\aff{1}, O.~Embreus\aff{1}, M.~Hoppe\aff{1}, G.~Papp\aff{2}  \and T.~F\"ul\"op\aff{1} } 

\affiliation{\aff{1}Department of Physics, Chalmers University of Technology,
 SE-41296 G\"{o}teborg, Sweden
\aff{2}Max Planck Institute for Plasma Physics, D-85748 Garching, Germany}

\begin{document}

\maketitle

\begin{abstract}
Integrated modelling of electron runaway requires computationally expensive kinetic models that are self-consistently coupled to the evolution of the background plasma parameters. The computational expense can be reduced by using parameterized runaway generation rates rather than solving the full kinetic problem. However, currently available generation rates neglect several important effects; in particular, they are not valid in the presence of partially ionized impurities. In this work, we construct a multilayer neural network for the Dreicer runaway generation rate which is trained on data obtained from kinetic simulations performed for a wide range of plasma parameters and impurities. The neural network accurately reproduces the Dreicer runaway generation rate obtained by the kinetic solver. By implementing it in a fluid runaway-electron modelling tool, we show that the improved generation rates lead to significant differences in the self-consistent runaway dynamics as compared to the results using the previously available formulas for the runaway generation rate.
 \end{abstract}
 
\section{Introduction}
If the electric field exceeds a critical field in plasmas, the
accelerating force on fast electrons overcomes the collisional
friction.  The electrons are then accelerated to relativistic energies -- they 
run away \citep{dreicer1959} -- until the electric force is balanced by
another mechanism such as synchrotron radiation. Runaway
acceleration occurs in solar flares \citep{Holman1985}, lightning
discharges \citep{Dwyer} and in magnetic fusion devices
\citep{helander2002,Breizman_2019}. In tokamaks, runaway electrons form when large electric  
fields are present: during startup of the discharge,
radio-frequency current drive or disruptions -- a sudden termination of a tokamak discharge. 
In high-current, reactor-scale machines, a disruption has the potential to convert a major
part of the plasma current to a relativistic runaway-electron beam. Such a beam would severely
damage plasma-facing components if it is not well controlled~\citep{Lehnen2015}. 

An accurate modelling capacity of runaway electrons is essential to
evaluate the different methods aimed at mitigating their effects.  To
describe the full evolution of the temperature, plasma composition, electric
field, magnetic equilibrium and runaway current, it would be necessary to simultaneously solve for the relativistic momentum-space dynamics, the magnetic field evolution (including breakup of magnetic surfaces) and the transport (including both collisional and turbulent processes). This is, however, unfeasible with currently available computational resources and will likely remain so in the foreseeable future.

Until simulations of the full plasma evolution during a disruption can
be realized, as an intermediate step, transport codes and equilibrium
solvers could be coupled with analytical formulas for runaway
generation rates.  This means that instead of evolving the full
runaway-electron distribution, only key quantities such as the runaway
 number density would be considered and computed from the
instantaneous electric field and background plasma parameters. 
In such fluid models, 
the runaway-electron density evolves by analytical generation rates describing Dreicer, hot-tail  and avalanche generation, as well as tritium decay and Compton
scattering of $\gamma$-rays (which can be emitted by the activated wall in the nuclear phase of tokamak operation).  This 
approach has been used in the past to gain insight into the runaway-electron
dynamics and electric-field diffusion; some examples are the {\sc go}
code \citep{Smith2006GO,gal,Feher,papp13effect} and the work by
\citet{MartinSolis2017}.

Although runaway fluid models are useful to understand runaway dynamics, present tools use  runaway generation rates that lack several important effects, including the effect of synchrotron radiation~\citep{Stahl2015},  bremsstrahlung~\citep{EmbreusBrems2016}, and screening effects in partially ionized plasmas~\citep{Hesslow}. 
The need for amendments is particularly pronounced in scenarios involving disruption mitigation by massive material injection. The injected impurity ions will radiate the thermal energy content of the plasma during the thermal quench, and will become weakly ionized in the resulting cold plasma, implying that the nuclei will be partially screened by the bound electrons in interactions with fast electrons.  Recent studies indicate substantial differences in the runaway dynamics compared to the fully ionized case: the effect of partial screening might give order of magnitude differences in both the Dreicer generation rate~\citep{HesslowJPP} and the avalanche multiplication factor~\citep{Hesslow_2019}.

While the avalanche growth rate calculation  \citep{RosenbluthPutvinski1997} was recently generalized to include the effect of partial screening and radiation reaction \citep{Hesslow_2019}, 
a similar generalization of the Dreicer generation rate calculation by \citet{connor} appears intractable. This is because the  
  Dreicer rate is exponentially sensitive to the plasma properties at near-thermal energies, where the collision frequencies have a complicated energy dependence in the presence of cold impurities. Instead, \citet{martinsolis1} suggested to replace the Dreicer field $E\sub{D}$ and the effective charge $Z\sub{eff}$ in the \citet{connor} formula, but otherwise keep the same form of the expression. This approach has however not been validated by solutions of the kinetic equation.

A different approach to improving the Dreicer generation rate is to use a large database of numerical solutions of the kinetic equation to train a  neural network. Neural networks have proved to be useful to fit complicated, high-dimensional results to multi-parameter models, which is the case here since the density of each impurity species, in combination with a wide range of plasma parameters, gives a large number of free parameters. Neural networks are  widely used in many areas of physics, including fusion plasma physics; see e.g.~\citet{Svensson_1999, Clayton_2013, Citrin2015,Boyer_2019}.

In this paper, we construct a neural network that determines the
Dreicer generation rate including the effect of partial screening as well
as an energy-dependent Coulomb logarithm.  After detailing the kinetic
model and its validity (\S~\ref{sec:model}), we use a neural network
to create a model of the Dreicer generation rate based on kinetic
simulations (\S~\ref{sec:NN}). The resulting network successfully
reproduces the runaway generation rates predicted by the original
kinetic solver, and can be directly implemented into integrated
models. As a proof of concept, we implement the network into the runaway fluid simulation tool
\textsc{go}, and demonstrate that partial screening has a
significant effect on runaway
generation~(\S~\ref{sec:application}). Finally, we discuss the
applications of the model as well as possible
improvements~(\S~\ref{sec:concl}).

\section{Kinetic model}
\label{sec:model}

The \cite{dreicer1959,dreicer1960} mechanism for runaway generation
 originates from  the interplay between collisional energy diffusion and electric-field acceleration.  
In a time-independent background plasma, the Dreicer mechanism causes a constant particle flow through any momentum-space boundary beyond the runaway separatrix\footnote{Above the second separatrix, where fast particles are decelerated due to for example radiation reaction losses, the particles will slow down again. This eventually creates a bump-on-tail~\citep{HirvijokiBump2015, DeckerBump2016, guo2017}, which prevents further runaway growth through the Dreicer mechanism, but the involved energies and timescales makes this effect irrelevant except for near the threshold $E\sub{c}$. An exception is with strong synchrotron radiation, temperatures of several keV, and an electric field approaching the effective critical electric field; in this case a steady-state Dreicer generation rate cannot clearly be determined and a full kinetic simulation may be necessary.}; this flow rate defines the  steady-state Dreicer generation rate, 
\begin{equation}
\gamma \equiv 
\frac{\rd n\sub{\sc re}}{\rd t}.
\end{equation} To be useful in runaway fluid modelling, the system must be well described by a quasi-steady-state approximation, so that $\gamma(E, T, n\sub{e}, \dots, t)\approx\gamma[E(t), T(t), n\sub{e}(t),\dots]$, where $E, T, n\sub{e}, \dots$ are the plasma parameters which influence the Dreicer generation rate in steady state, such as the electric field $E$, the electron temperature $T$, and the electron density $n\sub{e}$. The quasi-steady-state generation rate thus requires sufficiently slowly varying parameters.  

The most accurate analytical treatment of the Dreicer  problem was carried out by \cite{connor}, who extended the results of~\cite{kruskal} to include relativistic effects, resulting in
\begin{align}
\gamma 
&= C \frac{n\sub{e}}{\tau\sub{ee}} \left(\frac{E}{E\sub{D}}\right)^{-\frac{3}{16}(1+Z\sub{eff})h} \exp\left[- \lambda\frac{E\sub{D}}{4E} - \sqrt{\eta\frac{(1+Z\sub{eff})E\sub{D}}{E}}\right], \label{eq:Dreicer}\\
\lambda &= 8\frac{E^2}{E\sub{c}^2}\left[1 - \frac{1}{2}\frac{E\sub{c}}{E} - \sqrt{1-\frac{E\sub{c}}{E}}\right], \nonumber \\
\eta &= \frac{1}{4}\frac{E^2}{E\sub{c}(E-E\sub{c})} \left[\frac{\pi}{2}-\arcsin\left(1-\frac{2E\sub{c}}{E}\right)\right]^2, \nonumber \\
h &= \frac{1}{3}\frac{1}{\frac{E}{E\sub{c}}-1}\left[\frac{E}{E\sub{c}}+2\left(\frac{E}{E_c}-2\right)\sqrt{\frac{E}{E-E\sub{c}}} - \frac{Z\sub{eff}-7}{Z\sub{eff}+1}\right]. \nonumber
\end{align}
Here, $E\sub{D} \tightequal n\sub{e} e^3 \ln{\Lambda_0}/(4 \pi \epsilon_0^2 T) $ is the Dreicer field, $E\sub{c}\tightequal E\sub{D} T/(m\sub{e} c^2) $ is the critical electric field,  
$Z\sub{eff}\tightequal\sum_i n_i Z_i^2 / n\sub{e}$ is the effective plasma charge, and $\tau\sub{ee}\tightequal 4\pi\varepsilon_0^2 m\sub{e}^2 v\sub{Te}^3/(n\sub{e} e^4\ln\Lambda_0 )$ is the thermal electron collision time, with the thermal speed $v\sub{Te} = \sqrt{2T/m\sub{e}}$, and  $\ln \Lambda_0$ is the Coulomb logarithm evaluated at the thermal speed (in the original work, $\ln \Lambda$ was assumed to be energy independent). 
The order-unity parameter $C$ is undetermined by~\citet{connor} but has been quantified in subsequent work to around $C \approx 1.0$ with our time normalization~\citep{kruskal,jayakumar1993}, and the parameters $\lambda$, $\eta$ and $h$ constitute the relativistic generalization of the generation rate; they approach unity as $E/E\sub{c} \rightarrow \infty$. 

\begin{figure}
  \begin{center}
\includegraphics[width=0.475\textwidth]{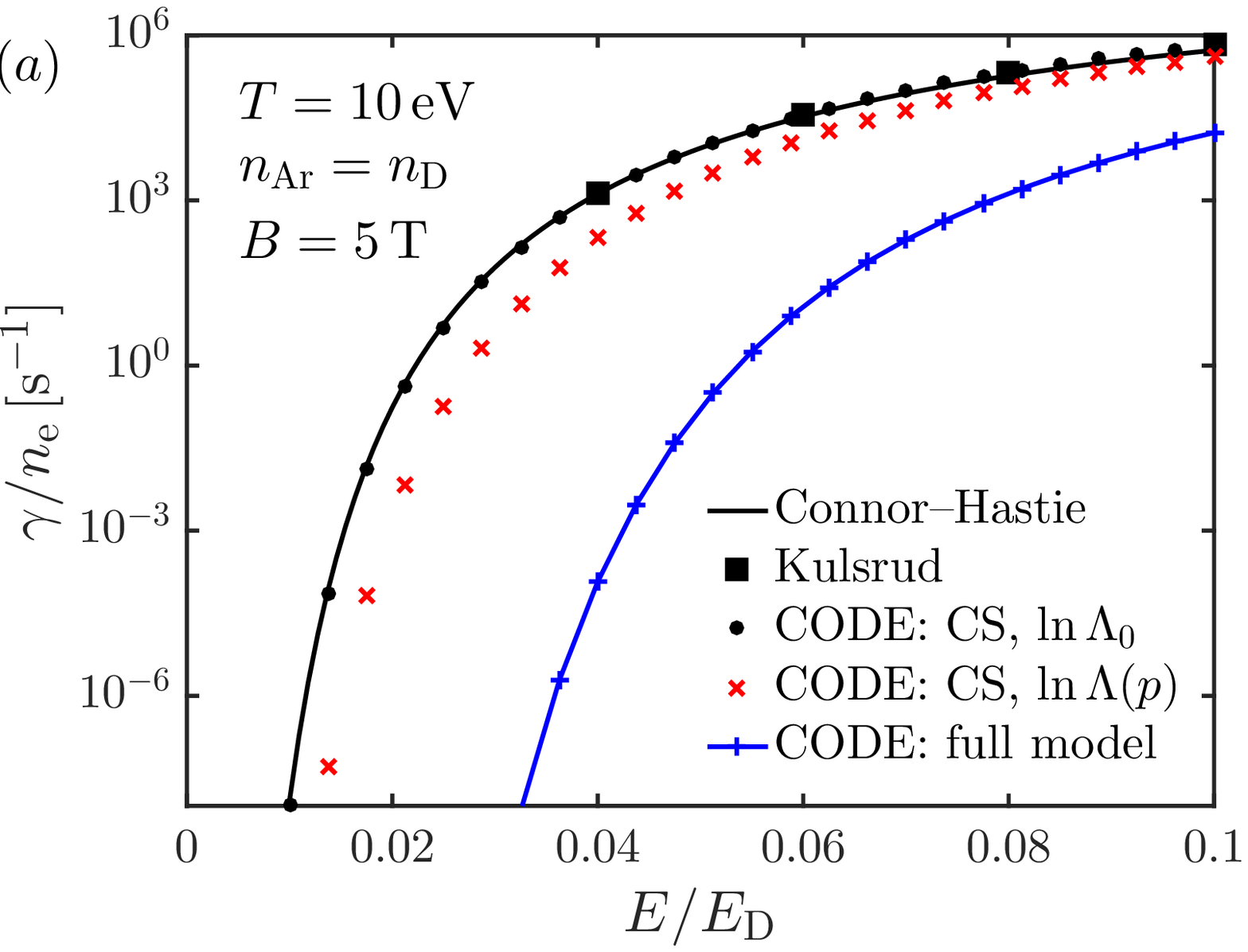} \hspace{1mm}
\includegraphics[width=0.475\textwidth]{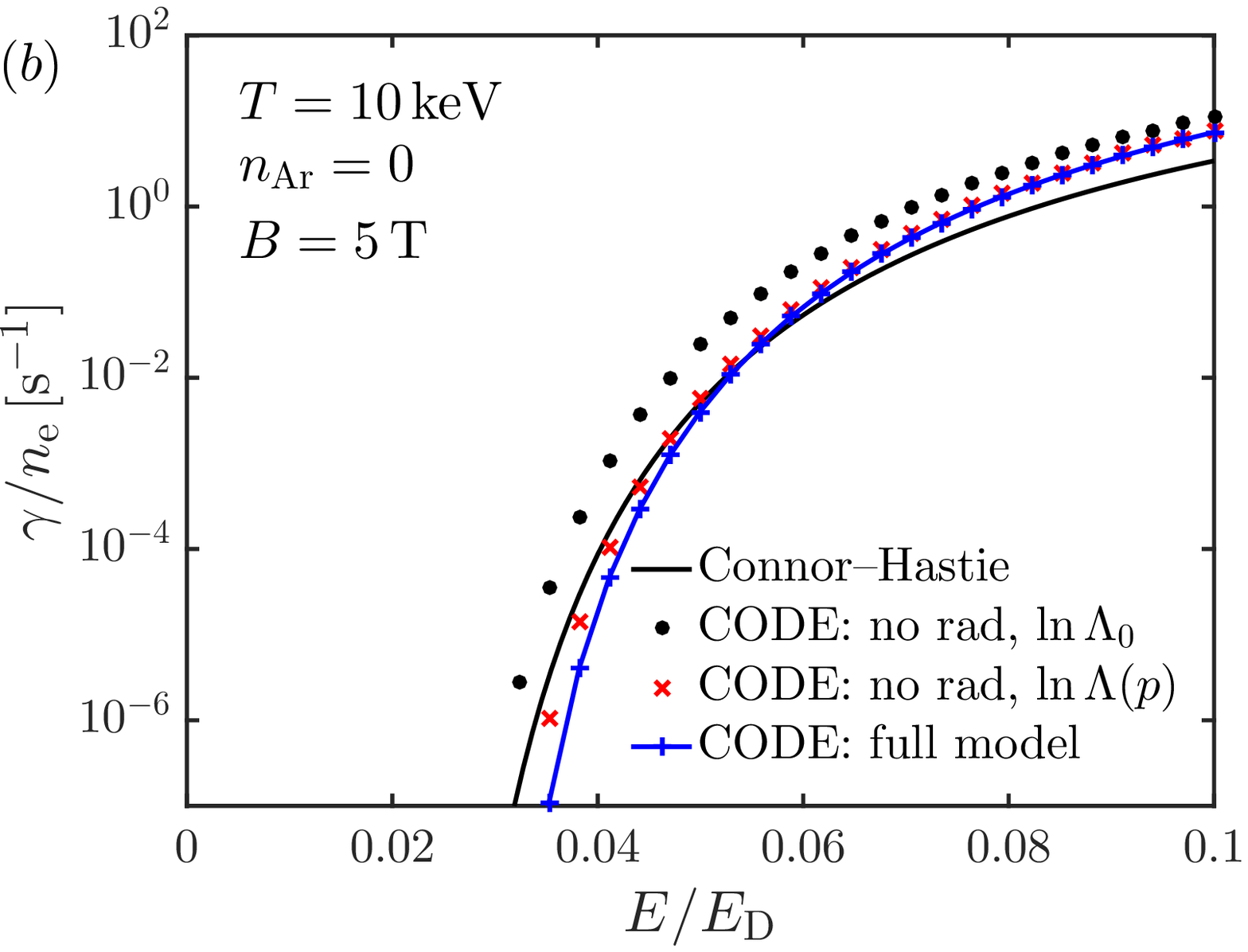}
\caption{\label{fig:Dreicer} Effect of partial screening, radiation reaction and an energy-dependent Coulomb logarithm on the Dreicer generation rate. Panel (\emph{a}) has singly ionized argon impurities with density $n\sub{Ar} = n\sub{D} = \unit[10^{20}]{m^{-3}}$ at a temperature of $T = \unit[10]{eV}$, whereas panel (\emph{b}) displays the effect of synchrotron and bremsstrahlung radiation reaction in a \unit[5]{T} magnetic field plasma with $n\sub{D} = \unit[10^{20}]{m^{-3}}$ and $T = \unit[10]{keV}$. In both (\emph{a}) and (\emph{b}), 
the solid black line shows the analytical formula 
from equation~\eqref{eq:Dreicer} with $C=1$, which was derived in the compeletely screened limit (denoted ``CS'' in the legend), using a constant Coulomb logarithm and neglecting the effect of radiation (denoted ``no rad'').  
Black dots and red crosses represent the simplified ideal plasma model where screening and radiation effects are ignored; 
black dots show \textsc{code} simulations using a constant Coulomb logarithm, whereas red crosses account for its energy dependence. The blue solid line with plus markers shows results that account for all kinetic effects.
For comparison, the black squares show the numerical results by \cite{kulsrud} in (\emph{a}), which were obtained in the nonrelativistic limit.}
\end{center}\end{figure}

Despite its apparent complexity, the generation rate in equation~\eqref{eq:Dreicer} neglects certain effects that are necessary for an accurate treatment of the problem. As illustrated in figure~\ref{fig:Dreicer}, the generation rate deviates significantly from equation~\eqref{eq:Dreicer} already when accounting for the energy dependence of the Coulomb logarithm $\ln \Lambda$, an effect which is most pronounced at the lower temperatures of figure~\ref{fig:Dreicer}$a$. By furthermore considering the effect of partial screening, the difference can reach several orders of magnitude. 

At high temperatures (figure~\ref{fig:Dreicer}$b$), the main deviation from the ideal behaviour is the energy dependence of the Coulomb logarithm. Since radiation reaction only becomes important at highly relativistic energies, these losses have a small effect on Dreicer generation except for where the critical momentum fulfils $p\sub{c} \gg 1$, which is obtained at near-critical electric fields if the temperature is high. As shown in figure~\ref{fig:Dreicer}$b$, the effect of synchrotron and bremsstrahlung radiation reaction is however modest even at relatively synchrotron-favouring parameters with $T = \unit[10]{keV}$, $B = \unit[5]{T}$ and $n\sub{D} = \unit[10^{20}]{m^{-3}}$.   
In these simulations, radiation reaction mainly affects the generation rate close to the effective critical field, which is consistent with previous work~\citep{Stahl2015}. 
We also note that Dreicer generation predominantly takes place at low temperatures during disruption scenarios, since the value of $E/E\sub{D} \sim j/(\sigma E\sub{D}) \sim 1/\sqrt{T}$ decreases with temperature at constant current density. For the remainder of this work, we therefore neglect the effect of radiation reaction on the Dreicer generation rate.

We also disregard the effect of toroidicity, which may have an appreciable effect on Dreicer generation off the magnetic axis: if the bounce time is much shorter than the detrapping time, the generation rate is reduced due to magnetic trapping~\citep{Nilsson2015}. Conversely, at high densities and electric fields $E \gg E\sub{c}$, which can be present during tokamak
disruptions, the Dreicer generation is approximately local (as in this work), but will be spatially non-uniform since the induced electric field decreases in magnitude with major radius~\citep{McDevitt_2019}.

The results in this work, including figure~\ref{fig:Dreicer}, were
obtained with the linearized Fokker-Planck solver {\sc code}
\citep{CODEPaper2014,Stahl2016}, which models a spatially homogeneous,
magnetized plasma. The test-particle collision operator in \textsc{code} is given in appendix~\ref{app:collop}. 
Since the collision operator is linearized, it is only valid for weak electric fields $E \ll E\sub{D}$. Otherwise, a major part of the thermal electron distribution runs away and a nonlinear Fokker-Planck equation, solved by for example {\sc norse} \citep{NORSE}, should be used. Conversely, the linear and nonlinear operators should agree at weak electric fields, and we verify in appendix~\ref{app:collop} that the test-particle collision operator in \textsc{code} gives a similar generation rate to \textsc{norse}.

\textsc{code} has a model of partial screening based on the Born
approximation~\citep{HesslowJPP}.  This is the most significant
limitation of this work, since the Born approximation is strictly valid only 
for electron speeds obeying $v/c \gg Z \alpha$, where
$\alpha\approx1/137$ is the fine-structure constant and $Z$ is the
atomic number. For argon and neon, the validity of the Born
approximation has been experimentally verified to extend beyond this requirement, down to kinetic energies of approximately \unit[1]{keV}~\citep{MottMassey}.
Below this threshold, the model for partial screening is
approximate, although it has been asymptotically matched to the correct behaviour as
$p\rightarrow 0$. 
 As the Dreicer generation rate is most sensitive to the
dynamics near the critical momentum given by
$p\sub{c} \gtrsim (E/E\sub{c}-1)^{-1/2}$, the model is only strictly
valid for $\unit[E/E\sub{D}]{(\%)} \lesssim T/(\unit[20]{eV}$), implying that the accuracy of our screening 
model is compromised for certain parameters. 

Notably, we found that the generalization of the generation rate suggested by \cite{MartinSolis2017} -- to replace $ Z\sub{eff}$ and $E\sub{D}$ by expressions involving the increased collision rates evaluated at an approximate value of the critical momentum $p\sub{c}$ -- gave poor agreement with kinetic simulations. A more involved amendment of the Dreicer generation rate~\eqref{eq:Dreicer} is therefore required, and would likely result in a significantly more involved analysis than was done by \cite{connor}. For this reason, we resorted to the use of a neural network, which will be described in the following section. 

\section{Neural network model for the Dreicer generation rate}
\label{sec:NN}

We used {\sc code} to determine the steady-state momentum-space distribution of the fast electrons, from which we determine the normalized generation rate 
\begin{equation}
\bar \gamma \equiv \frac{\tau\sub{ee}}{n\sub{e}}\gamma.
\end{equation}
In order to minimize the computational cost of the simulations, \textsc{code} was used in the steady-state mode, as described by \cite{CODEPaper2014} and detailed in appendix~\ref{app:flowVelocity}.

\textsc{code} simulations were performed for a large number of points randomly  sampled in the region described in table
\ref{tab:scan}, where $n_Z$ is the impurity ion density, and $n\sub{D}$ is the density of deuterium (or other hydrogen species;  the isotope does not affect the generation rate). 
For each ionization state of argon and neon, i.e.~${\rm Ar}^{+n}$ for $n=0\cdots 18$, and ${\rm Ne}^{+m}$ for $m=0\cdots 10$ (one at a time), 8000 points were sampled.
Additionally, 10000 points were sampled without any impurities ($n_Z = 0$), and 10000 points with a mix of different impurities with total density $n_Z$. The maximum temperature was set to either 
\unit[20]{keV} or twice the mean excitation energy, $2I_Z$; the latter for cases with a single impurity species. 
This is because a given charge state does not typically occur at higher temperatures. For example, $I_{\rm Ar^{+}}\tightequal\unit[219.4]{eV}$ \citep{sauer2015}, at which temperature argon would already be  multiply ionized in equilibrium. Moreover, in our model for partial screening, the enhancement of the slowing-down collision frequency starts to extend into the thermal population at such temperatures~\citep{HesslowJPP}, and the validity of the screening model starts to become questionable. 
\begin{table}
  \begin{center}
    \def~{\hphantom{0}}
    \begin{tabular}{ lccc }
      Parameter & Range & Scale\\[3pt]
      $n\sub{D}$ [$\unit{m^{-3}}$] & $10^{18} -  10^{21}$ & logarithmic \\
      $n_Z/n\sub{D}$ & $0 - 10$ & linear \\
      $E/E\sub{D}$ & $0.01 - 0.13$ & linear\\
      $T$ & $1 \mathrm{eV} - \unit[20]{keV} \textrm{ or } 2I_Z$ & logarithmic\\
    \end{tabular}
    \caption{Parameters used in the {\sc code} simulations, their range and how they were sampled.}
    \label{tab:scan}
  \end{center}
\end{table}

Rather than using the full set of impurity ion densities as input to the neural network, we use six derived parameters:
\begin{align}
\label{eq:derivedParams}
\ln n\sub{e}, \quad 
Z_\mathrm{eff}, \quad 
\frac{n\sub{e}}{n_\text{tot}}, \quad 
\sum_i \frac{n_i}{n_\text{tot}} (Z_i^2-Z_{0,i}^2), \quad
\sum_i \frac{n_i}{n_\text{tot}} Z_{0,i}Z_i, \quad  \mbox{and} \quad 
\sum_i \frac{n_i}{n_\text{tot}} \frac{Z_{0,i}}{Z_i}.
\end{align}
Here, $Z_i$ and $Z_{0,i}$ are the atomic number and charge number of species $i$, respectively, and $n_\mathrm{tot} = \sum_i Z_i n_i$ is the total density of free and bound electrons. 
These derived parameters were chosen to both include the relevant parameters in the completely screened limit ($n\sub{e}$ and $ 
Z_\mathrm{eff}$) parameters that naturally appear in the partially screened collision frequencies ($\nofrac{n\sub{e}}{n_\text{tot}}$  
 and $\sum_i (\nofrac{n_i}{n_\text{tot}}) (Z_i^2-Z_{0,i}^2)$), as well as the last two parameters in equation~\eqref{eq:derivedParams}, which vary significantly with plasma composition. 
This reduced the required size of the network, and allowed it to generalize better to other impurity species (for example, the neural network could accurately predict the Dreicer generation rate with carbon impurities although it was not trained for these). 
Accordingly, the input data was composed of 8 parameters: six density-related inputs, the normalized electric field and the natural logarithm of the temperature. This input, combined with the output $\bar \gamma$,
was randomly split into a training and validation set with a 4:1 ratio.

The neural network\footnote{The neural network is available at \url{https://github.com/unnerfelt/dreicer-nn}. } is designed as a multilayer perceptron described by the following series of matrix multiplications and function applications:
\begin{equation}
  \ln \bar \gamma = \JPPmatrix{W}_5\tanh\left\{\JPPmatrix{W}_4  \cdots  \tanh(\JPPmatrix{W}_1 \vec{x} + \vec{b}_1) + \cdots + \vec{b}_4 \right\} + \vec{b}_5. 
\end{equation}
Here, $\vec{x}$ is the input vector described above, the matrices $\JPPmatrix{W}_{(k)}$ describe the weights
between the layers, and the vectors $\vec{b}_{(k)}$ are the biases.
The activation function, $\tanh$, is applied element-wise to the
four hidden layers which were of size 20 (i.e.\ $\JPPmatrix{W}_1\in \mathbb{R}^{20\times 8}$, $\{ \JPPmatrix{W}_2, \JPPmatrix{W}_3, \JPPmatrix{W}_4\} \in \mathbb{R}^{20\times 20}$,  and $\JPPmatrix{W}_5 \in \mathbb{R}^{1\times 20} $). 
 To determine the optimized weight and bias values, we used the Adam algorithm~\citep{AdamAlgorithm}, which is a stochastic gradient descent method.
The training of the neural network was implemented in Python using the library PyTorch~\citep{pyTorch}. The validation set was used to determine when the optimization was sufficiently converged and to avoid overfitting.

\begin{figure}
  \centering
\includegraphics{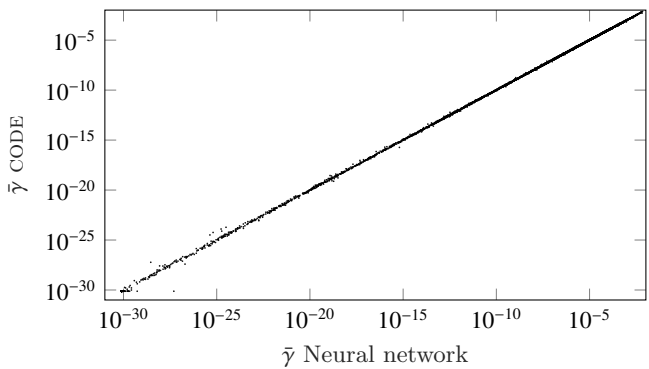}
\caption{\label{NN}Comparison between normalized runaway generation rates obtained from {\sc code} and those from the neural network regression.}
\end{figure}

Due to feedback between the current and the electric field, an order-unity error in the Dreicer generation rate will have a marginal impact on the maximum runaway current at the end of a current quench. This implies that some errors can be accepted, although the dominating factors must be correctly modelled to capture the final runaway-electron current. A comparison between the regression neural network and {\sc code} outputs for a set of 6500 points (different from both the validation and training set) is shown in figure~\ref{NN}.  We found that \unit[99.6]{\%} of the neural network predictions were within a factor of two of the correct value of $\gamma$, and the mean absolute
logarithmic (base 10) error was 0.0283. 
This indicates that the training data provided sufficient coverage of the parameter space. 
The evaluation time of the neural network is approximately five orders of magnitude faster than a steady-state {\sc code} simulation. The difference is even larger if the neural network is compared to a full time-dependent simulation with varying parameters and simultaneously accounting for avalanche multiplication, which is the type of simulation the generation rates can replace when used in integrated modelling. 

\begin{figure}
  \centering
\includegraphics{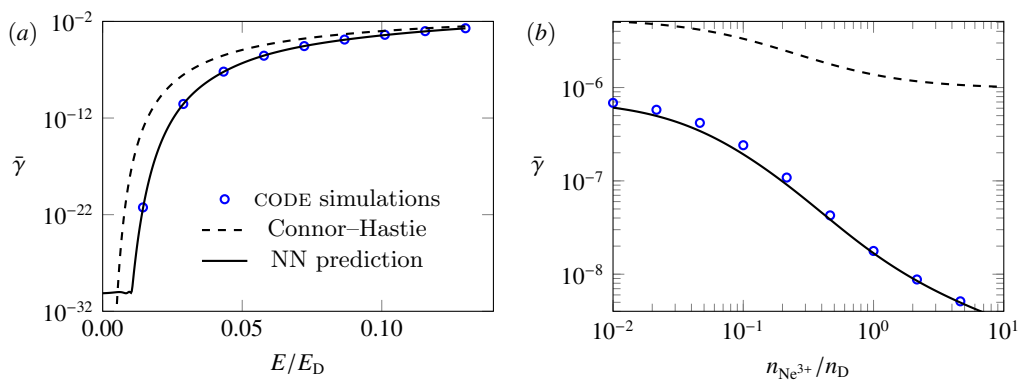}
\caption{\label{fits}Comparison between the normalized generation rate obtained by the neural network (solid lines), simulations with the {\sc code} Fokker-Planck solver (blue dots), and the analytical formula from  equation~\eqref{eq:Dreicer} with $C=1$ (dashed lines). The temperature was set to \unit[10]{eV}, and $n\sub{D}=\unit[10^{20}]{m^{-3}}$. In ($a$) $n_{\mathrm{Ne}^{3+}}/n\sub{D} = 1$, and in ($b$) $E / E_\mathrm{D} = 0.04$.}
\end{figure}

The typical quality of the fits is shown in figure~\ref{fits} where we display the normalized generation rate for a temperature of $T=\unit[10]{eV}$ and deuterium density $n\sub{D}=\unit[10^{20}]{m^{-3}}$. Figure~\ref{fits}$a$ shows the runaway generation rate as a function of normalized electric field calculated by the neural network, together with {\sc code} simulations, showing excellent agreement. For reference, the analytical formula~\eqref{eq:Dreicer} is also included (dashed line). 
Figure~\ref{fits}$b$ shows the generation rate as  a function of the density of triply ionized neon normalized to the density of hydrogen. Again, the agreement between the simulations and neural network is excellent, whereas the disagreement with the analytical expression is substantial.

\section{Application in runaway current modelling}
\label{sec:application} To demonstrate the impact of the modified Dreicer generation rates, we use the neural network
in a self-consistent simulation of the electric field and current
profile evolution performed by the {\sc go} numerical tool \citep{Smith2006GO}. Instead of
modelling the velocity space dynamics for the electrons in the runaway
region, {\sc go}  considers only their total density $n\sub{\sc re}$. It
solves the coupled equations for the runaway generation and resistive diffusion of the electric field, which in elongated plasmas reads \citep{Fulopupcoming}
\begin{equation}
  \frac{1+\kappa^{-2}}{2r}\frac{\partial}{\partial r}\left(r \frac{\partial E}{\partial r}\right)=\mu_0 \frac{\partial}{\partial t}(\sigma_\parallel E+ n\sub{\sc re} e c),\label{eq:ind}
\end{equation} where $E$ is the electric field, $\sigma_\parallel$ is the Spitzer conductivity with a neoclassical correction and $\kappa$ is the elongation. 

The model employed in \textsc{go} has several
limitations, e.g.\ it neglects radial losses due to magnetic
perturbations and coupling to the coils 
(i.e. \textsc{go} has a perfectly conducting wall at radius $r = b$, which gives an electric-field boundary condition at the plasma edge $r = a$ by matching to the vacuum solution).
Therefore, the numerical results are not expected to match the experimental values exactly, and the following examples are shown only as an illustration of the magnitude of the effect expected in an experimentally relevant scenario.

As an example of a scenario where the effect of partially ionized atoms is expected to be important, we consider JET discharge \#79423, in which a disruption was triggered with injection of $7.4\cdot 10^{20}$ argon atoms and a runaway-electron plateau of $\simeq\unit[590]{kA}$ was observed. The experimental parameters and the details of the discharge are described by \citet{papp13effect}. The pre-disruption parameters in the simulation were: major radius $R=\unit[3]{m}$;  minor radius $a=\unit[0.88]{m}$; radius of the conducting wall $b=\unit[1.3]{m}$; initial plasma current $I_\text{p}=\unit[1.93]{MA}$; magnetic field on axis $B=\unit[2]{T}$; elongation $\kappa=1.3$; density $n\sub{e}(r)=n_0 (1-1.27 \cdot r^2)^{0.43}$, with
$n_0=\unit[2.59\cdot 10^{19}]{m^{-3}}$; and temperature
$T(r)=T_0 (1-1.03 \cdot r^2)^2$ with $T_0=\unit[2.17]{keV}$ and $r$ is the normalized radial distance from the magnetic axis.

In the {\sc go} simulations,  we only included  Dreicer and avalanche runaway sources (no hot-tail generation). For the avalanche growth rate, we use the formula derived in \cite{Hesslow_2019}, which has been shown to give accurate results compared to numerical solutions of the kinetic equation.
The temperature evolution was taken from the experiment, but the post-disruption measurements exhibit a high degree of noise. 
 To correct for this artefact, we set the central temperature to $T_0=\unit[20]{eV}$ after the thermal quench, which gives a current quench time that agrees with the experiment. 
During the thermal quench, the ionization of the impurities was determined by calculating the density of each charge state for every ion species ($n^k_{Z_i}, k=0 \cdots Z_i$):
\begin{equation*}
    \frac{\rd n_{Z_i}^k}{\rd t}=  n\sub{e} \left( I_{k-1} n_{Z_i}^{k-1} - (I_k+ R_k) n_{Z_i}^{k} +  R_{k+1} n_{Z_i}^{k+1} \right),
\end{equation*}
where $I_{k}$ denotes the electron impact ionization rate for the
$k$-th charge state and $R_k$ is the radiative recombination rate, 
which were both taken from the ADAS database~\citep{ADAS}.  
After the disruption, the exact amount of argon 
was unknown as the assimilation is highly uncertain, and we
 therefore scanned over the argon density, which was assumed to be uniformly spread throughout the plasma. It is reasonable to assume
that in connection with the disruption some additional carbon will
penetrate into the plasma, and therefore we present results with both
\unit[0]{\%} and \unit[20]{\%} additional carbon.

\begin{figure} 
  \begin{center}
\includegraphics[width=0.7\textwidth]{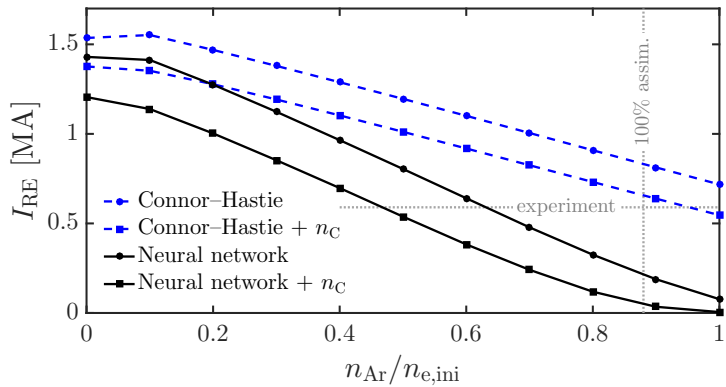}
\caption{Plateau runaway current as a function of argon density normalized to the initial electron density. Blue dashed lines are simulations with the \cite{connor} formula and solid black lines are simulations with the neural network. Simulations with squares include \unit[20]{\%} additional carbon, $n\sub{C} = 0.2 n\sub{e,ini}$. The experimental value ($\unit[590]{kA}$) is shown with dotted horisontal line, and  the vertical dotted line shows the value for \unit[100]{\%} assimilation, corresponding to $n\sub{Ar}/n\sub{e,ini}=0.88$.\label{fig:JET}}\end{center}\end{figure}

Figure \ref{fig:JET} shows the plateau runaway current given by {\sc
  go} as a function of argon density, using the \cite{connor}
analytical formula or the neural network, with or without additional carbon. 
With the analytical formula, argon densities corresponding to more than \unit[100]{\%} assimilation are required to match the experimental value, whereas lower   assimilation is sufficient with the neural network.  

\begin{figure}
  \begin{center}  
    \includegraphics[ height=0.30\textwidth]{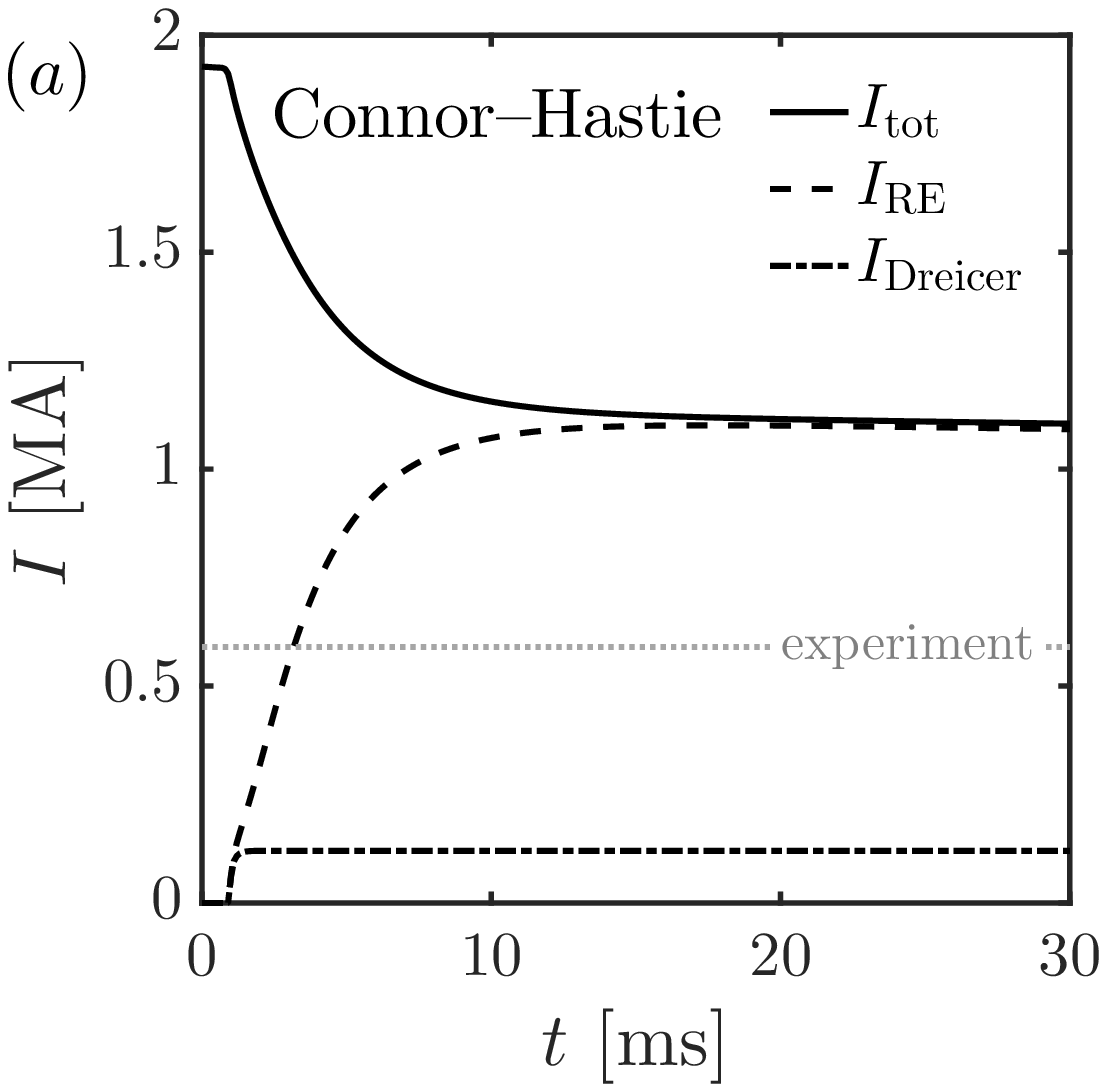}\hspace{0mm}
  \includegraphics[height=0.30\textwidth]{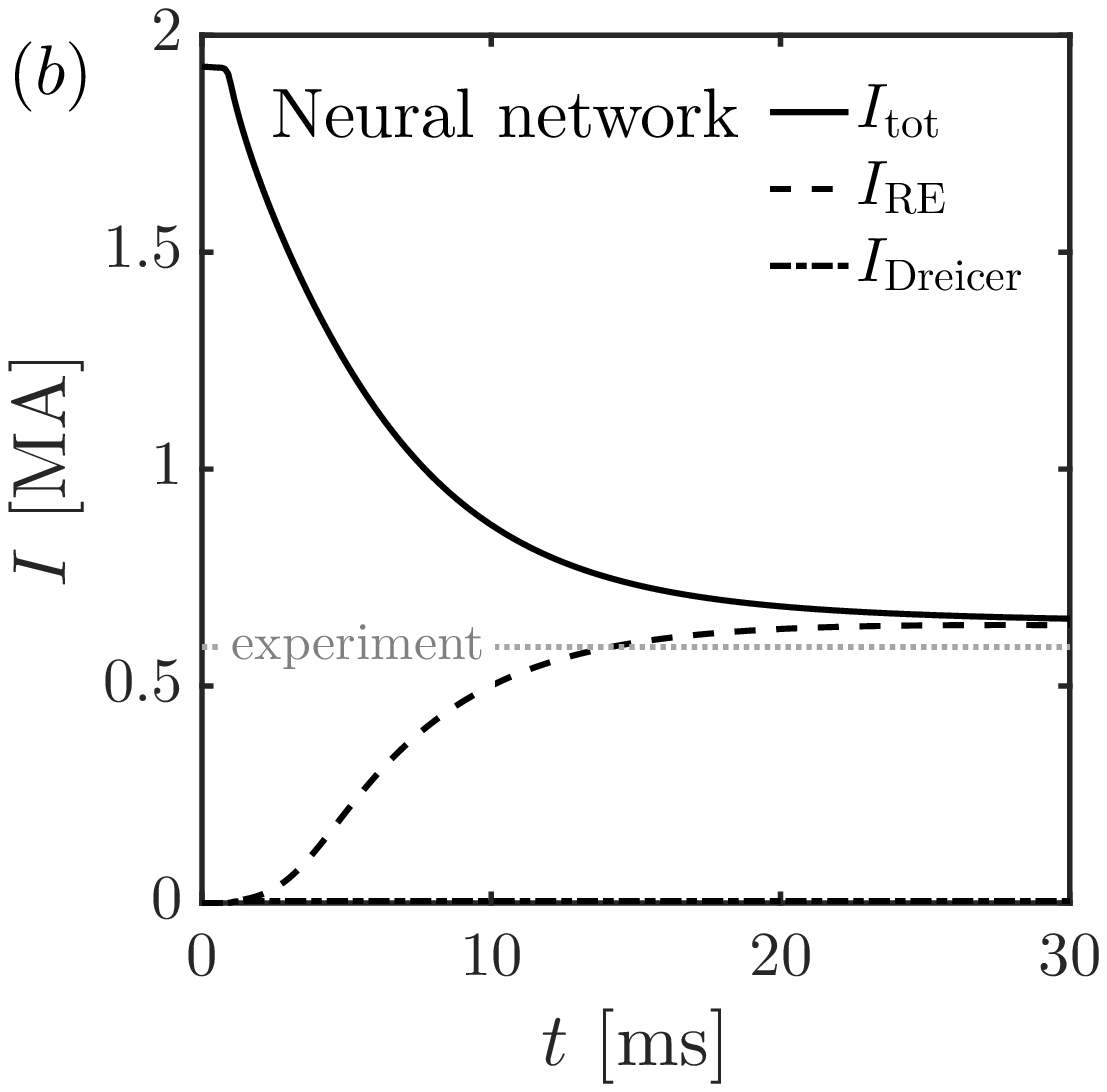}\hspace{0mm}
  \includegraphics[height=0.3\textwidth]{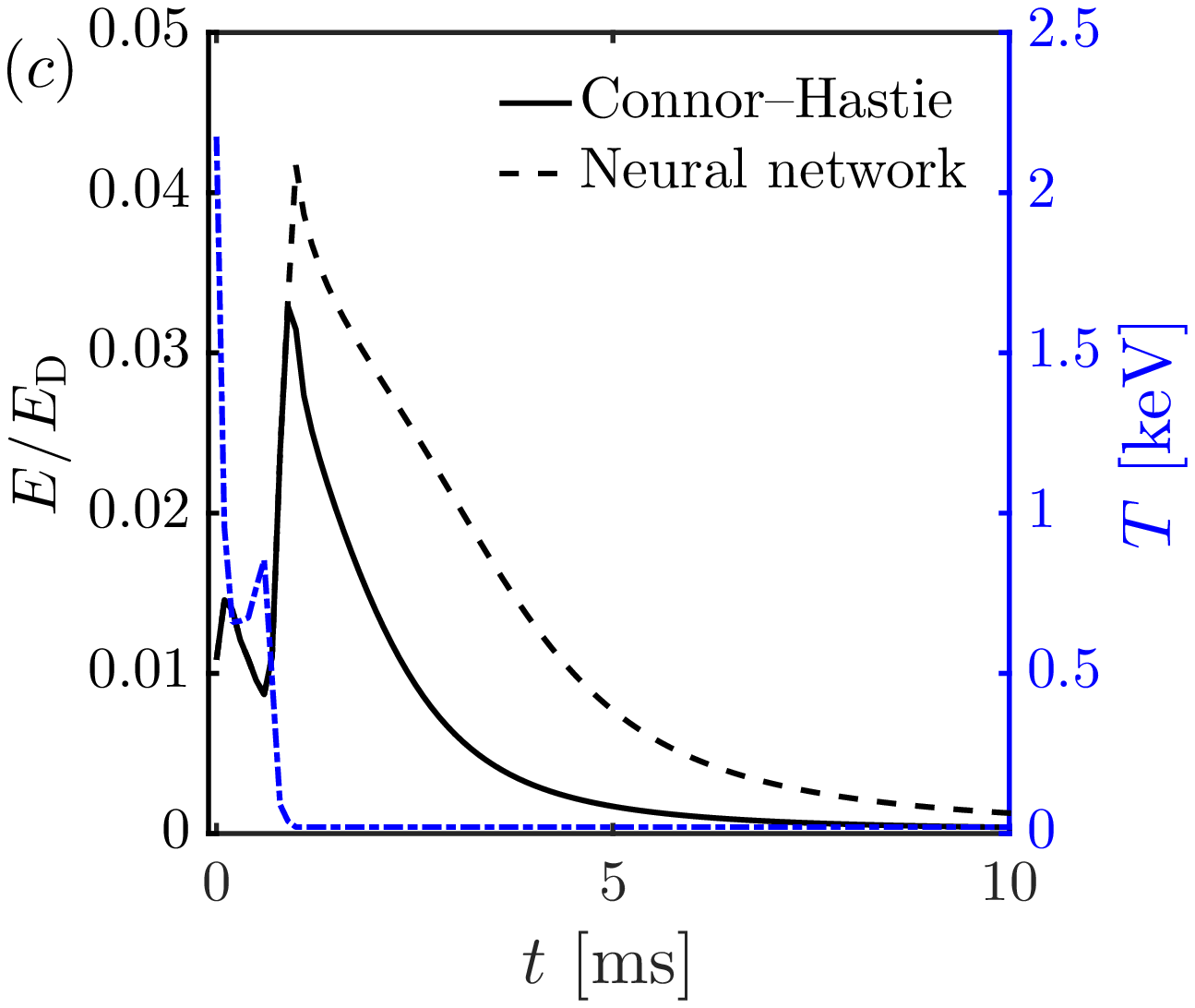}
\caption{Time evolution of the plasma
current for the case when the current approximately matches the experimental value ($n\sub{Ar}/n\sub{e,ini}=0.6$, $n_{\rm C}=0$) with ($a$) the \cite{connor} formula  and ($b$) the neural network. The Dreicer current is shown by dotted lines, and dashed lines show the total runaway current (Dreicer+avalanche).
In panel ($c$), the left $y$-axis shows the induced electric field (Connor--Hastie formula in solid line, neural network in dashed line), and the right $y$-axis shows the temperature in dash-dotted blue line. Both quantities were evaluated at the magnetic axis. Note the shorter time scale in ($c$) compared to ($a$) and ($b$).
\label{fig:JETcurrent}}\end{center}\end{figure}

Figure \ref{fig:JETcurrent}$a, b$ shows the time evolution of the plasma current for a case when the current density nearly matches the experimental value ($n\sub{Ar}/n\sub{e,ini}=0.6$, $n\sub{C}=0$) comparing the analytical prediction with the neural network. The Dreicer generation mainly occurs in a short interval around \unit[1]{ms} (although the Dreicer seed is barely visible in figure \ref{fig:JETcurrent}$b$). As  shown figure \ref{fig:JETcurrent}$c$, this time period coincides with the largest values of $E/E\sub{D}$,  
which is expected as the Dreicer generation rate is highly sensitive to this normalized electric field. 
Comparing figures \ref{fig:JETcurrent}$a$ and \ref{fig:JETcurrent}$b$, 
the neural network predicts a significantly reduced Dreicer seed, which is only partially compensated by the increased avalanche multiplication in the self-consistent electric field in figure \ref{fig:JETcurrent}$c$. The result is an order-unity reduction in the plateau runaway current.  

\section{Discussion and conclusions}
\label{sec:concl}
Runaway acceleration of particles occurring in plasmas with strong
electric fields has been studied for more than a century, but only
recently has it become possible to perform kinetic simulations in complex scenarios.  However, coupling such kinetic calculations to a self-consistent simulation of
the evolution of the background plasma parameters still poses a
significant computational challenge. It is therefore useful to have runaway generation rates that can be used in so-called runaway fluid models, i.e.\ integrated runaway simulations where generation rates are used to evolve the runaway current instead of solving the full kinetic problem. 

In this work, we presented a neural network that determines the steady-state Dreicer runaway generation rate accounting for collisions with partially ionized atoms. The network was trained on a large number of kinetic simulations and gives accurate results for the Dreicer runaway generation rate in plasmas consisting of hydrogen isotopes, neon and argon. This tool can therefore be used for rapid evaluation of such generation rates in any
laboratory, space or astrophysical plasma where runaway electrons are
produced. In particular, the tool should be valuable in simulations of tokamak disruptions involving massive material injection, in which case partially ionized impurities are expected to play an important role for the dynamics. Together with previously developed generation rates for avalanche, hot-tail, tritium decay and Compton scattering, the Dreicer generation rate neural network offers an improved runaway fluid model. 

In their present form, runaway fluid models have some limitations. Most notably, these models typically relate the number density generation rate to the associated current density by approximating the mean parallel runaway velocity $\langle v_\parallel \rangle\tightapprox c$. This assumption is expected to be more accurate in 
avalanche-dominated scenarios, where the runaway seed population carries a negligible part of the current. In such scenarios, the runaway number density is amplified through the avalanche mechanism while the average speed approaches the speed of light. Conversely, if a significant part of the current is converted into a runaway current through the Dreicer or hot-tail mechanisms, the assumption $\langle v_\parallel \rangle\tightapprox c$ may significantly overestimate the runaway current. As the mean velocity evolves in time during runaway generation in such scenarios, $\langle v_\parallel \rangle$ cannot be determined by steady-state simulations like those performed here, but would require time-dependent modelling including the history of the electric field. Nevertheless, this generation rate tool offers an improvement of the fluid models already without accounting for such effects, especially because runaway generation is generally avalanche dominated in the present and future large tokamaks that carry multi-MA plasma currents.

As an illustration, we implemented this model in 
\textsc{go} \citep{Smith2006GO}, and 
presented numerical solutions of the coupled equations of runaway
generation and electric-field diffusion in a JET-like disruptive scenario. 
In this scenario, we observed that the plateau runaway current was significantly reduced when using the neural network instead of the Connor--Hastie formula, which demonstrates the need to account for partially ionized atoms for realistic modelling of Dreicer generation. The neural network presented here can therefore be useful to improve runaway-electron modelling in tokamak simulation codes such as \textsc{astra} \citep{ASTRA_runaway}, \textsc{jorek} \citep{JOREK_Bandaru}, and \textsc{ets} \citep{ETS_Pokol}.  

\section*{Acknowledgements}
The authors are grateful to  S~Newton for fruitful
discussions.  This work was supported by the Swedish Research Council
(Dnr.~2018-03911), the Knut and Alice Wallenberg Foundation and the
European Research Council (ERC-2014-CoG grant 647121).  This work was supported by the EUROfusion - Theory and Advanced Simulation Coordination (E-TASC). This work has been carried out within the framework of the EUROfusion Consortium and has received funding from the Euratom research and training programme 2014-2018 and 2019-2020 under grant agreement No 633053. The views and opinions expressed herein do not necessarily reflect those of the European Commission.
\appendix
\section{Collision operator used in the Fokker-Planck simulations}
\label{app:collop}

The kinetic equation solved by \textsc{code} \citep{CODEPaper2014,Stahl2016} describes a magnetized, homogeneous plasma: 
\begin{align}
 \frac{\partial \fe}{\partial t} + \underbrace{ e E 
 \!\left(\xi \frac{\partial \fe }{\partial p} + \frac{1\!-\!\xi^2}{p}\frac{\partial \fe}{\partial \xi}\right)}_{\rm electric\, field} 
= \underbrace{\vphantom{\frac{\partial}{\partial p}}
C\sub{FP} 
}_{\rm collisions}&+ \underbrace {\vphantom{\frac{\partial}{\partial p}}
  C_{\rm br}-
\frac{\partial}{\partial \vec{p}}\!\cdot\! \left(\vec{F}_{\rm syn} \fe\right)
}_{\rm radiation \, reaction} + S
 \label{eq:CODEeq},
\end{align} 
where $\fe$ is the electron distribution function, $E$ is the component of the electric field that is antiparallel to the magnetic field $\vec{B}$, and $\xi = \vec{p}\cdot\vec{B}/(pB)$ is the cosine of the pitch-angle. Collisions are modelled by the Fokker--Planck collision operator (as this work focuses on Dreicer generation, a large-angle collision operator describing avalanche generation is not included). Radiation losses are modelled by $C_{\rm br}$ (the bremsstrahlung collision operator) and $\vec{F}_{\rm syn}$ (the synchrotron radiation reaction force). 
The Fokker-Planck solver {\sc code} includes several options for the collision operator. In this work we use the test-particle collision operator given by \cite{BraamsKarney1989} and \cite{PikeRose}, with corrections for partial screening according to \cite{HesslowJPP}:
\begin{align}
C\sub{FP}&=\frac{1}{p^2}\frac{\partial}{\partial p}\left[p^3\left(\nu\sub{s}\fe + \frac{1}{2}\nu_{\parallel}p\frac{\partial \fe}{\partial p}\right)\right]+\frac{1}{2}\nu\sub{D}\frac{\partial}{\partial\xi}\left[\left(1-\xi^2\right)\frac{\partial \fe}{\partial\xi}\right],
\end{align}
where the slowing-down, parallel and deflection frequencies are given by, respectively,  
\begin{equation}
\label{eq:collfreqs}
\left. \begin{array}{ll}  
\nu\sub{s}  
= \displaystyle\frac{1}{\tau}\frac{\gamma^2}{ \bar{p}^3}\frac{1}{\ln\Lambda_0}\left(\ln\Lambda\supp{ee}\frac{\bar p^2}{\gamma^2}\Psi\sub{s}(\bar{p},\Theta) + h(\bar p)  \right), \\[8pt]
\nu_{\parallel}
= \displaystyle\frac{1}{\tau}\frac{2\gamma \Theta}{\bar{p}^3}\Psi\sub{s}(\bar{p},\Theta),\\[8pt]
\nu\sub{D}
= \displaystyle\frac{1}{\tau}\frac{\gamma}{\bar p^3}\frac{1}{\ln \Lambda_0}\left(\ln \Lambda\supp{ee}  \frac{2\bar p}{\gamma}\Psi\sub{D}(\bar{p},\Theta)+ \ln \Lambda\supp{ei} Z\sub{eff} +  g(\bar{p})\right).
\end{array}\right\}
\end{equation}
Here, $\bar p \tightequal p /(m\sub{e} c)$ is the normalized momentum, $\Theta = T/(m\sub{e}c^2)$, and $\tau \tightequal \nofrac{4 \pi \varepsilon_0^2 m\sub{e}^2 c^3}{(n\sub{e} e^4 \ln \Lambda_0)}$ is a relativistic collision time. The energy-dependent Coulomb logarithm is modelled by matching the thermal Coulomb logarithm $\ln\Lambda_0 = 14.9-0.5 \ln(n\sub{e}[\unit[10^{20}]{m^{-3}}])+\ln
(T[\unit{keV}])$ from \cite{wesson} and the high-energy formula from \cite{SolodovBetti}, according to
\begin{equation}
\left. \begin{array}{ll}  
\ln \Lambda^{\rm ee} = \ln\Lambda_0 +
\displaystyle\frac{1}{k}\ln\left(1+\left[2(\gamma\!-\!1)/\bar p^2\sub{Te}\right]^{k/2} \right),
\\[8pt]
 \ln \Lambda^{\rm ei} = \ln\Lambda_0 +
\displaystyle\frac{1}{k} \ln\left[1+\left(2 \bar p/\bar p\sub{Te}\right)^k\right],
\end{array}\right\}
\end{equation}
 where $\bar p\sub{Te} = \sqrt{2T/(m\sub{e} c^2)}$ is the normalized thermal momentum and the parameter $k=5$ is chosen to give a smooth transition. 
The contributions from the fully ionized plasma are captured by the functions 
\begin{equation}
\left. \begin{array}{ll}  
\Psi\sub{s}(\bar p,\Theta)=\displaystyle\frac{\gamma^2\Psi_1-\Theta\Psi_0+(\Theta\gamma-1)\bar p e^{-\gamma/\Theta}}{\bar p^2K_2(1/\Theta)},\\[8pt]
\Psi\sub{D}(\bar p,\Theta)=\displaystyle\frac{
\big(\bar p^2\gamma^2+\Theta^2\big)\Psi_0+\Theta\big(2\bar p^4-1\big)\Psi_1+\gamma\Theta\big[1+\Theta\big(2\bar p^2-1\big)\big]\bar pe^{-\gamma/\Theta}
}{2\gamma \bar p^3K_2(1/\Theta)},
\end{array}\right\}
\end{equation}
with $\Psi_0 = \int_0^{\bar p} \exp\big[{-}\sqrt{1+s^2}/\Theta\big]\big/\big(\!\sqrt{1+s^2}\big)\, \rd s$, $\Psi_1 = \int_0^{\bar p} \exp\big[{-}\sqrt{1+s^2}/\Theta\big]\rd s$, and $K_2$ is the second-order modified Bessel function of the second kind.
Finally, the partial screening corrections are 
\begin{equation}
\left. \begin{array}{ll}  
h(\bar p) = \sum_i \displaystyle\frac{n_i}{n_e}N_{{\rm e},i}\left\{\frac{1}{5}\ln\bigg[1+\bigg(\frac{\bar p \sqrt{\gamma-1}}{I_i/m_e c^2}\bigg)^5\bigg]-\beta^2 \right\} ,
\\[8pt]
g(\bar p) = \sum_i \displaystyle\frac{n_i}{n\sub{e}} \left\{\frac{2}{3}\left(Z_i^2-Z_{0,i}^2\right)\ln\left[(\bar p \bar a_i)^{3/2}+1\right] -\frac{2}{3} \frac{N_{{\rm e},i}^2 (\bar p \bar a_i)^{3/2}}{(\bar p \bar a_i)^{3/2}+1}\right\},
\end{array}\right\}
\end{equation}
where $Z_{0,i}$ is the charge number, $Z_i$ is the atomic number and $N_{{\rm e},i} = Z_i\!-\!Z_{0,i}$ is the number of bound electrons of the nucleus for species $i$.
The length parameter $\bar a_i$ is given in \cite{HesslowJPP} and the mean excitation energy $I_i$ in \cite{sauer2015}. 

The collision operator given above is a linearized relativistic
test-particle operator, i.e.\ the field-particle part of the operator
is neglected. To assess the discrepancy due to the
combined effect of the field-particle operator and nonlinear effects, 
we compared the runaway generation rates computed with {\sc code}, using
the test-particle operator given above, and simulations with the fully
nonlinear kinetic equation solver {\sc norse} \citep{NORSE}, which uses the collision
operator derived by \cite{BraamsKarney1989}. The simulations were performed at temperatures in the range $\unit[10]{eV}{-}\unit[10]{keV}$, and for electric fields $E/E\sub{D}\tightapprox \unit[1{-}3.5]{\%}$, corresponding to $E/E\sub{c}\tightapprox 2{-}2700$. In all simulations, we find agreement between \textsc{code} and \textsc{norse}  simulations within a factor of two, which is enough to capture the order of magnitude of the Dreicer seed. Below $\unit[1]{keV}$, where $E/E\sub{c} \gg 1$, the two tools agreed within \unit[15]{\%}. In contrast, when the temperature increases beyond $\unit[1]{keV}$, the differences were up to \unit[50]{\%}. This is because $E/E\sub{c}$ decreases with temperature at constant $E/E\sub{D}$, and since the Dreicer generation rate is dramatically sensitive to the electric field at near-critical values, even a negligible difference in the electric field can lead to a significant difference in the generation rate. In other words, the observed differences may be smaller than any reasonable uncertainty in the electric field. 
Given the orders-of-magnitude variation of the Dreicer generation rate with electric field, the errors here are deemed acceptable and sufficient to capture the magnitude of the Dreicer seed.

\section{Calculation of the steady-state runaway generation rate}
 \label{app:flowVelocity}

To rapidly calculate the Dreicer generation rate, {\sc code} can be used in steady-state mode as described by \cite{CODEPaper2014}. Equation~\eqref{eq:CODEeq} has the form 
\begin{equation}
\frac{\partial \fe}{\partial t} + M \fe = S,
\end{equation}
which implies that the steady-state distribution can be obtained by $\fe = M^{-1}S$, where $S$ is a source term that cancels the constant outflow of particles due to the Dreicer mechanism. The steady-state generation rate $\gamma$ can then be obtained by integrating equation~\eqref{eq:CODEeq} over $2 \pi  \int_{p_b}^\infty p^2 \rd p \int_{-1}^1 (\cdot) \rd \xi$, where $p\sub{b}$ is taken so that the source $S$ vanishes for $p\geq p\sub{b}$. If radiation reaction is neglected, this yields  
\begin{equation}
\gamma = -4 \pi \left[ 
-\frac{1}{3} e E p^2 f_1 + 
p^4\frac{1}{2}\nu_\parallel \frac{\partial f_0}{\partial p}
 +  p^3 \nu\sub{s} f_0 \right]_{p = p\sub{b}},
 \label{eq:gammaFlow}
\end{equation}
where $f_L = (2L+1)2^{-1}\int_{-1}^1 \fe P_L(\xi) \rd \xi$ is the $L$-th Legendre mode of the steady-state distribution.  
This generation rate is insensitive to the exact shape of $S$ as long as it is limited to the bulk and does not affect the runaway population~\citep{CODEPaper2014}; accordingly, the source is taken to be $S = \alpha e^{-p/p\sub{Te}^2}$, with the magnitude $\alpha$ determined from the normalization of $\fe$. If equation~\eqref{eq:gammaFlow} is numerically well resolved, the generation rate is also independent of $p\sub{b}$ as long as the source $S$ vanishes for $p\geq p\sub{b}$.
Here, we took the average value of $\gamma$ over $p\sub{b}/p\sub{Te} \in[10,50]$. The lower limit was raised if there were large variations within the interval, which may occur due to floating point errors when evaluating \eqref{eq:gammaFlow} for weak electric fields. 

\bibliographystyle{jpp}

\clearpage

\bibliography{references}

\begin{thebibliography}{48}
\expandafter\ifx\csname natexlab\endcsname\relax\def\natexlab#1{#1}\fi
\def\au#1{#1} \def\ed#1{#1} \def\yr#1{#1}\def\at#1{#1}\def\jt#1{\textit{#1}}
  \def\bt#1{#1}\def\bvol#1{\textbf{#1}} \def\vol#1{#1} \def\pg#1{#1}
  \def\publ#1{#1}\def\arxiv#1{#1}\def\org#1{#1}\def\st#1{\textit{#1}}

\bibitem[Bandaru {\em et~al.\/}(2019)Bandaru, Hoelzl, Artola, Papp \&
  Huijsmans]{JOREK_Bandaru}
{\sc \au{Bandaru, V.}, \au{Hoelzl, M.}, \au{Artola, F.~J.}, \au{Papp, G.} \&
  \au{Huijsmans, G. T.~A.}} \yr{2019}  \at{Simulating the nonlinear interaction
  of relativistic electrons and tokamak plasma instabilities: Implementation
  and validation of a fluid model}.  \jt{Phys. Rev. E}  \bvol{99},
  \pg{063317}.

\bibitem[Boyer {\em et~al.\/}(2019)Boyer, Kaye \& Erickson]{Boyer_2019}
{\sc \au{Boyer, M.}, \au{Kaye, S.} \& \au{Erickson, K.}} \yr{2019}
  \at{Real-time capable modeling of neutral beam injection on {NSTX}-u using
  neural networks}.  \jt{Nuclear Fusion}  \bvol{59}~(5),  \pg{056008}.

\bibitem[Braams \& Karney(1989)]{BraamsKarney1989}
{\sc \au{Braams, B.~J.} \& \au{Karney, C. F.~F.}} \yr{1989}  \at{Conductivity
  of a relativistic plasma}.  \jt{Physics of Fluids {B}: Plasma Physics}
  \bvol{1}~(7),  \pg{1355}.

\bibitem[Breizman {\em et~al.\/}(2019)Breizman, Aleynikov, Hollmann \&
  Lehnen]{Breizman_2019}
{\sc \au{Breizman, B.~N.}, \au{Aleynikov, P.}, \au{Hollmann, E.~M.} \&
  \au{Lehnen, M.}} \yr{2019}  \at{Physics of runaway electrons in tokamaks}.
  \jt{Nuclear Fusion}  \bvol{59}~(8),  \pg{083001}.

\bibitem[Citrin {\em et~al.\/}(2015)Citrin, Breton, Felici, Imbeaux, Aniel,
  Artaud, Baiocchi, Bourdelle, Camenen \& Garcia]{Citrin2015}
{\sc \au{Citrin, J.}, \au{Breton, S.}, \au{Felici, F.}, \au{Imbeaux, F.},
  \au{Aniel, T.}, \au{Artaud, J.}, \au{Baiocchi, B.}, \au{Bourdelle, C.},
  \au{Camenen, Y.} \& \au{Garcia, J.}} \yr{2015}  \at{Real-time capable first
  principle based modelling of tokamak turbulent transport}.  \jt{Nuclear
  Fusion}  \bvol{55}~(9),  \pg{092001}.

\bibitem[Clayton {\em et~al.\/}(2013)Clayton, Tritz, Stutman, Bell, Diallo,
  LeBlanc \& Podest{\`{a}}]{Clayton_2013}
{\sc \au{Clayton, D.~J.}, \au{Tritz, K.}, \au{Stutman, D.}, \au{Bell, R.~E.},
  \au{Diallo, A.}, \au{LeBlanc, B.~P.} \& \au{Podest{\`{a}}, M.}} \yr{2013}
  \at{Electron temperature profile reconstructions from multi-energy {SXR}
  measurements using neural networks}.  \jt{Plasma Physics and Controlled
  Fusion}  \bvol{55}~(9),  \pg{095015}.

\bibitem[Connor \& Hastie(1975)]{connor}
{\sc \au{Connor, J.} \& \au{Hastie, R.}} \yr{1975}  \at{Relativistic
  limitations on runaway electrons}.  \jt{Nuclear Fusion}  \bvol{15},
  \pg{415}.

\bibitem[Decker {\em et~al.\/}(2016)Decker, Hirvijoki, Embr\'eus, Peysson,
  Stahl, Pusztai \& F\"ul\"op]{DeckerBump2016}
{\sc \au{Decker, J.}, \au{Hirvijoki, E.}, \au{Embr\'eus, O.}, \au{Peysson, Y.},
  \au{Stahl, A.}, \au{Pusztai, I.} \& \au{F\"ul\"op, T.}} \yr{2016}
  \at{Numerical characterization of bump formation in the runaway electron
  tail}.  \jt{Plasma Physics and Controlled Fusion}  \bvol{58}~(2),
  \pg{025016}.

\bibitem[Dreicer(1959)]{dreicer1959}
{\sc \au{Dreicer, H.}} \yr{1959}  \at{Electron and ion runaway in a fully
  ionized gas. {I}}.  \jt{Phys. Rev.}  \bvol{115},  \pg{238}.

\bibitem[Dreicer(1960)]{dreicer1960}
{\sc \au{Dreicer, H.}} \yr{1960}  \at{Electron and ion runaway in a fully
  ionized gas. {II}}.  \jt{Phys. Rev.}  \bvol{117},  \pg{329}.

\bibitem[Dwyer(2007)]{Dwyer}
{\sc \au{Dwyer, J.~R.}} \yr{2007}  \at{Relativistic breakdown in planetary
  atmospheres}.  \jt{Physics of Plasmas}  \bvol{14},  \pg{042901}.

\bibitem[Embr\'eus {\em et~al.\/}(2016)Embr\'eus, Stahl \&
  F\"ul\"op]{EmbreusBrems2016}
{\sc \au{Embr\'eus, O.}, \au{Stahl, A.} \& \au{F\"ul\"op, T.}} \yr{2016}
  \at{Effect of bremsstrahlung radiation emission on fast electrons in
  plasmas}.  \jt{New Journal of Physics}  \bvol{18}~(9),  \pg{093023}.

\bibitem[Fable {\em et~al.\/}(2016)Fable, Pautasso, Lehnen, Dux, Bernert,
  Mlynek \& {the ASDEX Upgrade Team}]{ASTRA_runaway}
{\sc \au{Fable, E.}, \au{Pautasso, G.}, \au{Lehnen, M.}, \au{Dux, R.},
  \au{Bernert, M.}, \au{Mlynek, A.} \& \au{{the ASDEX Upgrade Team}}} \yr{2016}
   \at{Transport simulations of the pre{\textendash}thermal{\textendash}quench
  phase in {ASDEX} upgrade massive gas injection experiments}.  \jt{Nuclear
  Fusion}  \bvol{56}~(2),  \pg{026012}.

\bibitem[Feh\'er {\em et~al.\/}(2011)Feh\'er, Smith, F\"ul\"op \& G\'al]{Feher}
{\sc \au{Feh\'er, T.}, \au{Smith, H.~M.}, \au{F\"ul\"op, T.} \& \au{G\'al, K.}}
  \yr{2011}  \at{Simulation of runaway electron generation during plasma
  shutdown by impurity injection in {ITER}}.  \jt{Plasma Physics and Controlled
  Fusion}  \bvol{53}~(3),  \pg{035014}.

\bibitem[F\"ul\"op {\em et~al.\/}(2019)F\"ul\"op, Helander, Vallhagen, Embreus,
  Svensson, Creely, Howard \& Rodriguez-Fernandez]{Fulopupcoming}
{\sc \au{F\"ul\"op, T.}, \au{Helander, P.}, \au{Vallhagen, O.}, \au{Embreus,
  O.}, \au{Svensson, P.}, \au{Creely, A.~J.}, \au{Howard, N.~T.} \&
  \au{Rodriguez-Fernandez, P.}} \yr{2019}  \at{Effect of plasma elongation on
  current dynamics during tokamak disruptions}.  \jt{Submitted to Journal of
  Plasma Physics},  \arxiv{arXiv: 1909.13707}.

\bibitem[G\'al {\em et~al.\/}(2008)G\'al, Feh\'er, Smith, F\"ul\"op \&
  Helander]{gal}
{\sc \au{G\'al, K.}, \au{Feh\'er, T.}, \au{Smith, H.~M.}, \au{F\"ul\"op, T.} \&
  \au{Helander, P.}} \yr{2008}  \at{Runaway electron generation during plasma
  shutdown by killer pellet injection}.  \jt{Plasma Phys. and Controlled
  Fusion}  \bvol{50},  \pg{055006}.

\bibitem[Guo {\em et~al.\/}(2017)Guo, McDevitt \& Tang]{guo2017}
{\sc \au{Guo, Z.}, \au{McDevitt, C.~J.} \& \au{Tang, X.-Z.}} \yr{2017}
  \at{Phase-space dynamics of runaway electrons in magnetic fields}.
  \jt{Plasma Physics and Controlled Fusion}  \bvol{59}~(4),  \pg{044003}.

\bibitem[Helander {\em et~al.\/}(2002)Helander, Eriksson \&
  Andersson]{helander2002}
{\sc \au{Helander, P.}, \au{Eriksson, L.-G.} \& \au{Andersson, F.}} \yr{2002}
  \at{Runaway acceleration during magnetic reconnection in tokamaks}.
  \jt{Plasma Physics and Controlled Fusion}  \bvol{44},  \pg{B247}.

\bibitem[Hesslow {\em et~al.\/}(2018)Hesslow, Embr\'eus, Hoppe, DuBois, Papp,
  Rahm \& F\"ul\"op]{HesslowJPP}
{\sc \au{Hesslow, L.}, \au{Embr\'eus, O.}, \au{Hoppe, M.}, \au{DuBois, T.~C.},
  \au{Papp, G.}, \au{Rahm, M.} \& \au{F\"ul\"op, T.}} \yr{2018}
  \at{Generalized collision operator for fast electrons interacting with
  partially ionized impurities}.  \jt{Journal of Plasma Physics}
  \bvol{84}~(6),  \pg{905840605}.

\bibitem[Hesslow {\em et~al.\/}(2017)Hesslow, Embr\'eus, Stahl, DuBois, Papp,
  Newton \& F\"ul\"op]{Hesslow}
{\sc \au{Hesslow, L.}, \au{Embr\'eus, O.}, \au{Stahl, A.}, \au{DuBois, T.~C.},
  \au{Papp, G.}, \au{Newton, S.~L.} \& \au{F\"ul\"op, T.}} \yr{2017}
  \at{Effect of partially screened nuclei on fast-electron dynamics}.
  \jt{Phys. Rev. Lett.}  \bvol{118},  \pg{255001}.

\bibitem[Hesslow {\em et~al.\/}(2019)Hesslow, Embr{\'{e}}us, Vallhagen \&
  F\"ul\"op]{Hesslow_2019}
{\sc \au{Hesslow, L.}, \au{Embr{\'{e}}us, O.}, \au{Vallhagen, O.} \&
  \au{F\"ul\"op, T.}} \yr{2019}  \at{Influence of massive material injection on
  avalanche runaway generation during tokamak disruptions}.  \jt{Nuclear
  Fusion}  \bvol{59}~(8),  \pg{084004}.

\bibitem[Hirvijoki {\em et~al.\/}(2015)Hirvijoki, Pusztai, Decker, Embr\'eus,
  Stahl \& F\"ul\"op]{HirvijokiBump2015}
{\sc \au{Hirvijoki, E.}, \au{Pusztai, I.}, \au{Decker, J.}, \au{Embr\'eus, O.},
  \au{Stahl, A.} \& \au{F\"ul\"op, T.}} \yr{2015}  \at{Radiation reaction
  induced non-monotonic features in runaway electron distributions}.
  \jt{Journal of Plasma Physics}  \bvol{81},  \pg{475810502}.

\bibitem[Holman(1985)]{Holman1985}
{\sc \au{Holman, G.~D.}} \yr{1985}  \at{Acceleration of runaway electrons and
  {J}oule heating in solar flares}.  \jt{Astrophysical Journal}  \bvol{293},
  \pg{584--594}.

\bibitem[Jayakumar {\em et~al.\/}(1993)Jayakumar, Fleischmann \&
  Zweben]{jayakumar1993}
{\sc \au{Jayakumar, R.}, \au{Fleischmann, H.} \& \au{Zweben, S.}} \yr{1993}
  \at{Collisional avalanche exponentiation of runaway electtrons in electrified
  plasmas}.  \jt{Physics Letters A}  \bvol{172},  \pg{447 -- 451}.

\bibitem[Kingma \& Ba(2014)]{AdamAlgorithm}
{\sc \au{Kingma, D.~P.} \& \au{Ba, J.}} \yr{2014}  \at{Adam: A method for
  stochastic optimization}.  \jt{arXiv preprint arXiv:1412.6980}.

\bibitem[Kruskal \& Bernstein(1962)]{kruskal}
{\sc \au{Kruskal, M.} \& \au{Bernstein, I.~B.}} \yr{1962}  \at{{Princeton
  Plasma Physics Lab}}.  \jt{Report no. MATT-Q-20}  \pg{p. 172}.

\bibitem[Kulsrud {\em et~al.\/}(1973)Kulsrud, Sun, Winsor \& Fallon]{kulsrud}
{\sc \au{Kulsrud, R.~M.}, \au{Sun, Y.-C.}, \au{Winsor, N.~K.} \& \au{Fallon,
  H.~A.}} \yr{1973}  \at{Runaway electrons in a plasma}.  \jt{Phys. Rev. Lett.}
   \bvol{31},  \pg{690}.

\bibitem[Landreman {\em et~al.\/}(2014)Landreman, Stahl \&
  F\"ul\"op]{CODEPaper2014}
{\sc \au{Landreman, M.}, \au{Stahl, A.} \& \au{F\"ul\"op, T.}} \yr{2014}
  \at{Numerical calculation of the runaway electron distribution function and
  associated synchrotron emission}.  \jt{Computer Physics Communications}
  \bvol{185}~(3),  \pg{847}.

\bibitem[Lehnen {\em et~al.\/}(2015)Lehnen, Aleynikova, Aleynikov, Campbell,
  Drewelow, Eidietis, Gasparyan, Granetz, Gribov, Hartmann, Hollmann, Izzo,
  Jachmich, Kim, Ko{\v{c}}an, Koslowski, Kovalenko, Kruezi, Loarte, Maruyama,
  Matthews, Parks, Pautasso, Pitts, Reux, Riccardo, Roccella, Snipes, Thornton
  \& de~Vries]{Lehnen2015}
{\sc \au{Lehnen, M.}, \au{Aleynikova, K.}, \au{Aleynikov, P.}, \au{Campbell,
  D.}, \au{Drewelow, P.}, \au{Eidietis, N.}, \au{Gasparyan, Y.}, \au{Granetz,
  R.}, \au{Gribov, Y.}, \au{Hartmann, N.}, \au{Hollmann, E.}, \au{Izzo, V.},
  \au{Jachmich, S.}, \au{Kim, S.-H.}, \au{Ko{\v{c}}an, M.}, \au{Koslowski, H.},
  \au{Kovalenko, D.}, \au{Kruezi, U.}, \au{Loarte, A.}, \au{Maruyama, S.},
  \au{Matthews, G.}, \au{Parks, P.}, \au{Pautasso, G.}, \au{Pitts, R.},
  \au{Reux, C.}, \au{Riccardo, V.}, \au{Roccella, R.}, \au{Snipes, J.},
  \au{Thornton, A.} \& \au{de~Vries, P.}} \yr{2015}  \at{Disruptions in {ITER}
  and strategies for their control and mitigation}.  \jt{Journal of Nuclear
  Materials}  \bvol{463},  \pg{39}.

\bibitem[Mart\'{\i}n-Sol\'{\i}s {\em et~al.\/}(2015)Mart\'{\i}n-Sol\'{\i}s,
  Loarte \& Lehnen]{martinsolis1}
{\sc \au{Mart\'{\i}n-Sol\'{\i}s, J.~R.}, \au{Loarte, A.} \& \au{Lehnen, M.}}
  \yr{2015}  \at{Runaway electron dynamics in tokamak plasmas with high
  impurity content}.  \jt{Physics of Plasmas}  \bvol{22},  \pg{092512}.

\bibitem[Mart{\'{\i}}n-Sol{\'{\i}}s {\em
  et~al.\/}(2017)Mart{\'{\i}}n-Sol{\'{\i}}s, Loarte \& Lehnen]{MartinSolis2017}
{\sc \au{Mart{\'{\i}}n-Sol{\'{\i}}s, J.~R.}, \au{Loarte, A.} \& \au{Lehnen,
  M.}} \yr{2017}  \at{Formation and termination of runaway beams in {ITER}
  disruptions}.  \jt{Nuclear Fusion}  \bvol{57}~(6),  \pg{066025}.

\bibitem[McDevitt \& Tang(2019)]{McDevitt_2019}
{\sc \au{McDevitt, C.~J.} \& \au{Tang, X.-Z.}} \yr{2019}  \at{Runaway electron
  generation in axisymmetric tokamak geometry}.  \jt{{EPL} (Europhysics
  Letters)}  \bvol{127}~(4),  \pg{45001}.

\bibitem[Mott \& Massey(1965)]{MottMassey}
{\sc \au{Mott, N.~F.} \& \au{Massey, H. S.~W.}} \yr{1965} {\em The theory of
  atomic collisions\/}.  \publ{Clarendon Press Oxford}.

\bibitem[Nilsson {\em et~al.\/}(2015)Nilsson, Decker, Peysson, Granetz,
  Saint-Laurent \& Vlainic]{Nilsson2015}
{\sc \au{Nilsson, E.}, \au{Decker, J.}, \au{Peysson, Y.}, \au{Granetz, R.},
  \au{Saint-Laurent, F.} \& \au{Vlainic, M.}} \at{ \yr{2015} } \jt{Plasma Phys.
  Controlled Fusion}  \bvol{57},  \pg{095006}.

\bibitem[Papp {\em et~al.\/}(2013)Papp, F\"ul\"op, Feh\'er, de~Vries, Riccardo,
  Reux, Lehnen, Kiptily, Plyusnin, Alper \& contributors]{papp13effect}
{\sc \au{Papp, G.}, \au{F\"ul\"op, T.}, \au{Feh\'er, T.}, \au{de~Vries, P.},
  \au{Riccardo, V.}, \au{Reux, C.}, \au{Lehnen, M.}, \au{Kiptily, V.},
  \au{Plyusnin, V.}, \au{Alper, B.} \& \au{contributors, J.~E.}} \yr{2013}
  \at{The effect of iter-like wall on runaway electron generation in jet}.
  \jt{Nuclear Fusion}  \bvol{53}~(12),  \pg{123017}.

\bibitem[Paszke {\em et~al.\/}(2017)Paszke, Gross, Chintala, Chanan, Yang,
  DeVito, Lin, Desmaison, Antiga \& Lerer]{pyTorch}
{\sc \au{Paszke, A.}, \au{Gross, S.}, \au{Chintala, S.}, \au{Chanan, G.},
  \au{Yang, E.}, \au{DeVito, Z.}, \au{Lin, Z.}, \au{Desmaison, A.}, \au{Antiga,
  L.} \& \au{Lerer, A.}} \yr{2017} Automatic differentiation in {PyTorch}.
  \bt{In {\em NIPS Autodiff Workshop\/}}.

\bibitem[Pike \& Rose(2014)]{PikeRose}
{\sc \au{Pike, O.~J.} \& \au{Rose, S.~J.}} \yr{2014}  \at{Dynamical friction in
  a relativistic plasma}.  \jt{Phys. Rev. E}  \bvol{89},  \pg{053107}.

\bibitem[Pokol {\em et~al.\/}(2019)Pokol, Olasz, Erdos, Papp, Aradi, Hoppe,
  Johnson, Ferreira, Coster, Peysson, Decker, Strand, Yadikin, Kalupin \& {the
  EUROfusion-IM Team}]{ETS_Pokol}
{\sc \au{Pokol, G.~I.}, \au{Olasz, S.}, \au{Erdos, B.}, \au{Papp, G.},
  \au{Aradi, M.}, \au{Hoppe, M.}, \au{Johnson, T.}, \au{Ferreira, J.},
  \au{Coster, D.}, \au{Peysson, Y.}, \au{Decker, J.}, \au{Strand, P.},
  \au{Yadikin, D.}, \au{Kalupin, D.} \& \au{{the EUROfusion-IM Team}}}
  \yr{2019}  \at{Runaway electron modelling in the self-consistent core
  {E}uropean {T}ransport {S}imulator}.  \jt{Nuclear Fusion}  \bvol{59}~(7),
  \pg{076024}.

\bibitem[Rosenbluth \& Putvinski(1997)]{RosenbluthPutvinski1997}
{\sc \au{Rosenbluth, M.} \& \au{Putvinski, S.}} \yr{1997}  \at{Theory for
  avalanche of runaway electrons in tokamaks}.  \jt{Nuclear Fusion}  \bvol{37},
   \pg{1355--1362}.

\bibitem[Sauer {\em et~al.\/}(2015)Sauer, Oddershede \& Sabin]{sauer2015}
{\sc \au{Sauer, S.~P.}, \au{Oddershede, J.} \& \au{Sabin, J.~R.}} \yr{2015}
  \at{Chapter three - the mean excitation energy of atomic ions}.  \bt{In {\em
  Concepts of Mathematical Physics in Chemistry: A Tribute to Frank E. Harris -
  Part A\/}},  \st{Advances in Quantum Chemistry},  \vol{vol.~71},  \pg{p.~29}.
   \publ{Academic Press}.

\bibitem[Smith {\em et~al.\/}(2006)Smith, Helander, Eriksson, Anderson, Lisak
  \& Andersson]{Smith2006GO}
{\sc \au{Smith, H.}, \au{Helander, P.}, \au{Eriksson, L.-G.}, \au{Anderson,
  D.}, \au{Lisak, M.} \& \au{Andersson, F.}} \yr{2006}  \at{Runaway electrons
  and the evolution of the plasma current in tokamak disruptions}.  \jt{Physics
  of Plasmas}  \bvol{13}~(10),  \pg{102502}.

\bibitem[Solodov \& Betti(2008)]{SolodovBetti}
{\sc \au{Solodov, A.~A.} \& \au{Betti, R.}} \yr{2008}  \at{Stopping power and
  range of energetic electrons in dense plasmas of fast-ignition fusion
  targets}.  \jt{Physics of Plasmas}  \bvol{15}~(4),  \pg{042707}.

\bibitem[Stahl {\em et~al.\/}(2016)Stahl, Embr\'eus, Papp, Landreman \&
  F\"ul\"op]{Stahl2016}
{\sc \au{Stahl, A.}, \au{Embr\'eus, O.}, \au{Papp, G.}, \au{Landreman, M.} \&
  \au{F\"ul\"op, T.}} \yr{2016}  \at{Kinetic modelling of runaway electrons in
  dynamic scenarios}.  \jt{Nuclear Fusion}  \bvol{56}~(11),  \pg{112009}.

\bibitem[Stahl {\em et~al.\/}(2015)Stahl, Hirvijoki, Decker, Embr\'eus \&
  F\"{u}l\"{o}p]{Stahl2015}
{\sc \au{Stahl, A.}, \au{Hirvijoki, E.}, \au{Decker, J.}, \au{Embr\'eus, O.} \&
  \au{F\"{u}l\"{o}p, T.}} \yr{2015}  \at{Effective critical electric field for
  runaway electron generation}.  \jt{Physical Review Letters}  \bvol{114},
  \pg{115002}.

\bibitem[Stahl {\em et~al.\/}(2017)Stahl, Landreman, Embr\'eus \&
  F\"ul\"op]{NORSE}
{\sc \au{Stahl, A.}, \au{Landreman, M.}, \au{Embr\'eus, O.} \& \au{F\"ul\"op,
  T.}} \yr{2017}  \at{{NORSE}: A solver for the relativistic non-linear
  {F}okker--{P}lanck equation for electrons in a homogeneous plasma}.
  \jt{Computer Physics Communications}  \bvol{212},  \pg{279}.

\bibitem[Summers(2004)]{ADAS}
{\sc \au{Summers, H.~P.}} \yr{2004} The {ADAS} user manual, version 2.6.
  \url{http://www.adas.ac.uk}.

\bibitem[Svensson {\em et~al.\/}(1999)Svensson, von Hellermann \&
  K\"onig]{Svensson_1999}
{\sc \au{Svensson, J.}, \au{von Hellermann, M.} \& \au{K\"onig, R. W.~T.}}
  \yr{1999}  \at{Analysis of {JET} charge exchange spectra using neural
  networks}.  \jt{Plasma Physics and Controlled Fusion}  \bvol{41}~(2),
  \pg{315--338}.

\bibitem[Wesson(2011)]{wesson}
{\sc \au{Wesson, J.}} \yr{2011} {\em Tokamaks\/}, 4th edn.  \publ{Oxford
  University Press}.

\end{thebibliography}
\end{document}